\documentclass[draft]{agujournal2019}
\usepackage{url} 
\usepackage{lineno}
\usepackage[inline]{trackchanges} 
\usepackage{amsmath}

\draftfalse

\journalname{Geophysical Research Letters}

\begin{document}

\title{Data Driven Air Entrainment Velocity Parameterization by Breaking Waves}

\authors{
        Xiaohui Zhou\affil{1,2}, 
        Anton S. Darmenov\affil{2}, and 
        Kianoosh Yousefi\affil{3}
}

\affiliation{1}{Earth System Science Interdisciplinary Center, University of Maryland, College Park, MD 20740, USA}
\affiliation{2}{NASA/GSFC Mail Code 610.1, Greenbelt, MD 20771}
\affiliation{3}{Department of Mechanical Engineering, University of Texas at Dallas, Richardson, TX 75080, USA}

\correspondingauthor{Xiaohui Zhou}{xiaohuizhou4work@gmail.com}



\begin{keypoints}
\item Physically interpretable ML parameterization of wave-breaking air-entrainment velocity trained on 43-yr WAVEWATCH III simulations.
\item ML-based Va parameterization markedly improves on wind-only and semi-bulk schemes using wind speed and significant wave height.
\item Widely available wind and wave predictors enable use across wave models, reanalyses, and observations without retraining.
\end{keypoints}

\begin{abstract}
Wave breaking injects turbulence and bubbles into the upper ocean, modulating air–sea exchange of momentum, heat, gases, and sea-spray aerosols. These fluxes depend nonlinearly on sea state but remain poorly represented in coupled atmosphere–wave–ocean models, where air-entrainment velocity ($V_a$) is often parameterized using wind speed or significant wave height alone. We develop a global machine-learning parameterization of $V_a$ trained on a 43-year WAVEWATCH III simulation that resolves the breaker-front distribution $\Lambda(c)$ and associated energetics. A multilayer perceptron with seven physically motivated predictors (wind speed, wave height, wave age, steepness, direction, and depth) reproduces spectral-reference $V_a$ with high skill. The model reduces longstanding biases in bulk formulas, notably overestimation in swell-dominated low latitudes and underestimation in storm tracks. Applied globally, the parameterization improves estimates of bubble-mediated CO$_2$ transfer velocity and sea-salt aerosol emission, reducing errors by roughly an order of magnitude. Validation against independent HiWinGS observations supports robust performance in deep-water, high-wind conditions.

\end{abstract}

\section*{Plain Language Summary}
Breaking waves play an important role in how the ocean exchanges gases, heat, and particles with the atmosphere. When waves break, they trap air and create bubbles that greatly increase gas exchange (such as $CO_2$ moving into the ocean) and produce tiny sea-spray particles. Most climate models cannot explicitly simulate these breaking events, so they rely on simplified relationships based only on wind speed. These simplified formulas often produce large errors.

In this study, we use machine learning trained on decades of global wave model simulations to create a new way to estimate how much air is mixed into the ocean and sea salt emitted by breaking waves. The new method captures how wave conditions, not just wind, affect breaking. It predicts air entrainment, gas transfer, and sea salt emissions more accurately than traditional formulas. This approach can be used to improve weather forecasts, air quality studies, and long-term climate simulations by representing the ocean’s whitecap processes much more realistically.

\section{Introduction}
Breaking surface waves exert a first-order control on air-sea exchanges of momentum, heat, and gases by generating intense near-surface turbulence, entraining air, and producing whitecaps and bubble plumes that enhance surface renewal and bubble-mediated transfer  \cite<e.g.,>{Thorpe1995, Melville1996, deane2002scale, Melville2016, ruth2022three, pelaez2024dynamics, deike2025universal, di2025air}. Although whitecaps cover only a small fraction of the ocean surface, they account for a disproportionate share of wave-energy dissipation and can dominate gas transfer under moderate-to-high winds \cite{Banner2000,Callaghan2008images,Sutherland2016,callaghan2012observed}. Because wave breaking is intermittent and strongly modulated by sea state (e.g., wind forcing, wave age, steepness, and swell), it remains challenging to represent consistently in coupled atmosphere--wave--ocean models and in large-scale diagnostics derived from reanalyses and wave hindcasts. As a result, many applications still rely on wind-speed-only (or semi-bulk) formulations that can miss substantial regional and regime-dependent variability associated with changes in wave development and spectral shape \cite{Deike2017,zhou2023sea}.

A physically grounded framework for describing wave breaking is provided by the breaking-crest distribution $\Lambda(c)$, the mean length of breaking crests moving with phase speed $c$ per unit surface area \cite{phillips1985spectral}. Moments of $\Lambda(c)$ connect breaking kinematics to energetics and enable sea-state-dependent estimates of dissipation and breaking intensity. Practical approaches to estimate $\Lambda(c)$ in spectral wave models have made it possible to evaluate global breaking statistics and their implications for air--sea exchange in a consistent, model-based manner \cite{Romero2012,Romero2019}. Building on this foundation, air-entrainment velocity $V_a$ provides a compact descriptor of breaking-driven air injection, linking wave-breaking statistics to bubble-mediated pathways that influence gas transfer and aerosol production \cite{Deike2017}. Recent implementations of $\Lambda(c)$-based diagnostics in WAVEWATCH~III (WW3) have enabled global estimates of sea-state-dependent air entrainment and associated bubble-mediated gas fluxes \cite{Romero2019,zhou2023sea}. However, explicitly computing $\Lambda(c)$ and derived breaking diagnostics within operational or climate-scale workflows can be computationally expensive, and simplified parameterizations based only on wind speed can produce systematic biases \cite{zhou2023sea}.

In parallel, machine learning (ML) has emerged as an effective tool for learning nonlinear mappings in geophysical systems when the target quantity is well-defined and physically interpretable, and when the predictors are widely available from models or observations. ML approaches have been used across Earth system science for parameterization, emulation, and downscaling, including wave and air-sea applications where the goal is to capture sea-state dependence at low computational cost \cite<e.g.,>{JamesOthers2018, ODonnchaOthers2018, ODonnchaOthers2019, LouOthers2021, ZhangOthers2022, SunOthers2022, DakarOthers2023, XuOthers2023}. For wave-breaking-related quantities, a key requirement is portability: the model should be driven by predictors that are routinely available (e.g., $U_{10}$, $H_s$, wave age, steepness, direction, and depth) and should generalize across basins and regimes without ad hoc tuning. This motivates a surrogate approach that retains the physics embedded in the spectral, $\Lambda(c)$-based reference while providing an efficient diagnostic that can be evaluated everywhere that standard wind and wave fields exist.

Here we develop a data-driven, computationally inexpensive surrogate for $V_a$ by training a multilayer perceptron (MLP) on a 43-year global WW3 hindcast in which $\Lambda(c)$ and associated breaking energetics are diagnosed. The model uses seven physically motivated predictors (significant wave height $H_s$, 10-m wind speed $U_{10}$, wind direction, wave age $c_p/U_{10}$, wave steepness $k_pH_s/2$, and water depth) to reproduce the spectral-reference $V_a$ with high fidelity. We evaluate performance globally using withheld years (2019--2022) and validate against independent observations from HiWinGS \cite{Brumer2017}, focusing on the ability to capture both the magnitude and variability of air entrainment across sea states. Finally, we demonstrate how improved sea-state-dependent $V_a$ propagates into two applications that are sensitive to breaking-driven bubbles: bubble-mediated gas transfer velocity and sea-salt aerosol emission, highlighting differences relative to wind-only and semi-bulk parameterizations.

The paper is organized as follows. Section~2 describes the WW3 simulations, predictor construction, and the MLP training. Section~3 evaluates the parameterization globally and against HiWinGS observations and quantifies implications for bubble-mediated gas transfer and sea-salt emission. Section~4 summarizes conclusions and limitations.

\section{Methodology}\label{section:Method}
\subsection{Numerical Modeling}
The breaking-crest distribution framework $\Lambda(c)$ is implemented in the ST4 source term of WAVEWATCH~III (WW3) \cite{ardhuin2010semiempirical}. We use WW3 v7.14 in a global configuration at $0.5^{\circ}$ resolution with 42 frequencies and 36 directions, following \citeA{zhou2023sea}. The model is forced by 3-hourly 10-m neutral winds $U_{10}$ from JRA55-do \cite{tsujino2018jra} and includes monthly varying sea ice from a coupled ocean--sea-ice simulation forced by JRA55-do fields \cite{adcroft2019gfdl}. The resolved wavenumber range ($k=0.0016$--$4~\mathrm{rad~m^{-1}}$) spans long and short wind waves and enables explicit diagnosis of wave-breaking statistics following \citeA{Romero2019}. The wave-spectrum simulation setup and validation are described in \citeA{zhou2023sea}.

From the simulated wave spectra $F(k,\theta)$ we diagnose $\Lambda(c)$ and compute air-entrainment velocity $V_a$ using \citeA{Deike2017}:
\begin{equation}
V_a = \tilde{B}\int S(k)^{3/2}\frac{c^3}{g}\Lambda(c)\,dc,
\end{equation}
where $S(k)$ is the saturation spectrum, $c$ is phase speed, and $\tilde{B}=0.1$ is a constant. The $\Lambda(c)$ distribution is linked to the WW3 dissipation source term via the wave-breaking model of \citeA{Romero2019}. We archive 3-hourly fields for 1980--2022, including $V_a$, significant wave height $H_s$, friction velocity $u_*$, wave age ($c_p/U_{10}$), and steepness ($k_pH_s/2$).

\subsection{Neural network model and training}
\begin{figure}[!htbp]
    \centering
    \includegraphics[width=0.9\linewidth]{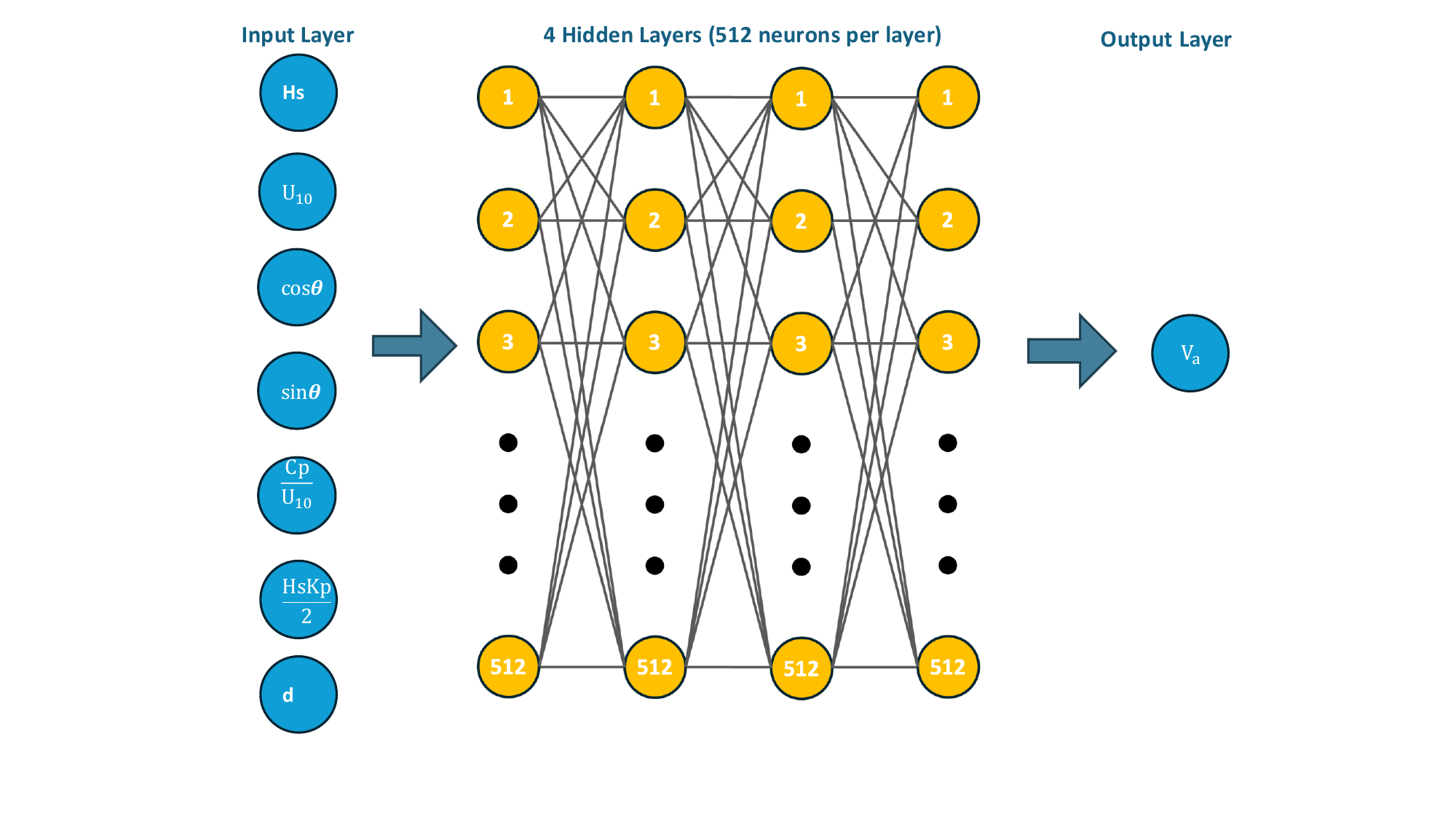}
    \caption{Schematic of the machine-learning framework used to parameterize the air-entrainment velocity ($V_a$). Seven physically based predictors are derived from WAVEWATCH III simulations: significant wave height ($H_s$), 10-m wind speed ($U_{10}$), cosine and sine of wind direction ($\cos\theta_w$, $\sin\theta_w$), wave age ($c_p/U_{10}$), wave steepness ($k_pH_s/2$), and water depth ($d$). These inputs are processed through a multilayer perceptron (MLP) with four hidden layers (512 neurons each, GELU activation, and $10\%$ dropout) to predict the target variable $V_a$. The architecture captures nonlinear relationships between wind forcing, sea-state parameters, and air entrainment, forming the basis for a global machine-learning parameterization of wave-breaking-induced air–sea fluxes.}
    \label{fig1}
\end{figure}

We train a multilayer perceptron (MLP) to predict $V_a$ from seven physically motivated predictors: $H_s$, $U_{10}$, wind direction $(\cos\theta,\sin\theta)$, wave age $c_p/U_{10}$, wave steepness $k_pH_s/2$, and water depth $d$ (Figure~\ref{fig1}). The MLP consists of four fully connected hidden layers with 512 neurons each and GELU activations, followed by a linear output layer. A dropout rate of 0.1 is applied in each hidden layer for regularization. 

We explored alternative depths (2, 4, and 6 layers) using a reduced diagnostic dataset; two-layer models underperformed and gains beyond four layers were marginal relative to added cost, so we adopt the four-layer configuration (Table~S1 in Supporting Information). The model is trained with Adam (learning rate $10^{-3}$, weight decay $10^{-4}$) to minimize mean squared error. Training uses mixed-precision, distributed data-parallel optimization across multiple GPUs with stochastic pointwise sampling from the multi-decadal global dataset, so an epoch corresponds to a fixed number of optimization steps rather than a complete pass through the data. Training and validation losses drop rapidly within the first 8-10 epochs and remain stable (Figure~S1), we therefore train the final model for 12 epochs.

We split the 1980-2022 dataset into training (1980-2014), validation (2015-2018), and testing (2019-2022), corresponding to 80\%, 10\%, and 10\% of the samples, respectively. Predicted $V_a$ from the MLP model was evaluated against withheld WW3 reference fields and independent HiWinGS observations \cite{Brumer2017} using $RMSE$, bias, and correlation. The model reproduces the expected dependence of $V_a$ on wind speed, wave age, and steepness implied by the $\Lambda(c)$ framework and performs consistently across developing and mature sea states.


\section{Results}
In this study, we treat the wave spectral simulated global air entrainment velocity ($V_a^{Spec}$) based on $\Lambda(c)$ as the ground truth. To compare the MLP model with existing parameterizations, we consider the works of \citeA{Deike2017} and \citeA{Liang2017} who suggested parameterizations of $V_a$ based on frictional velocity $u_*$, phase speed $C_p$ and significant wave height $H_s$: 
\begin{displaymath}
    V_a^{Semi} = V_a(C_p,u_*, H_s) = a_1 C_p(u_*/\sqrt{gH_s})^{b_1},
\end{displaymath}
and a simplified parameterization based on 10-meter wind speed $U_{10}$: 
\begin{displaymath}
    V_a^{U_{10}} = V_a(U_{10}) = a_2(U_{10}-c_2)^{b_2}.
\end{displaymath}
We use the same training dataset described in Section \ref{section:Method} to fit the above two relationships. The best fits obtained in this study are: $V_a(u_*, H_s) = 2.4\times 10^{-3}C_p(u_*/\sqrt{gH_s})^{2.48}$ and $V_a(U_{10}) = 5.5\times 10^{-6}(U_{10}-2.35)^{1.05}$. Note that in the following discussion, the units of $V_a$ are changed from $m/s$ to $cm/hour$.

\subsection{Global air entrainment from 2019 to 2022}
\begin{figure}[!htbp]
    \noindent\includegraphics[width=\textwidth]{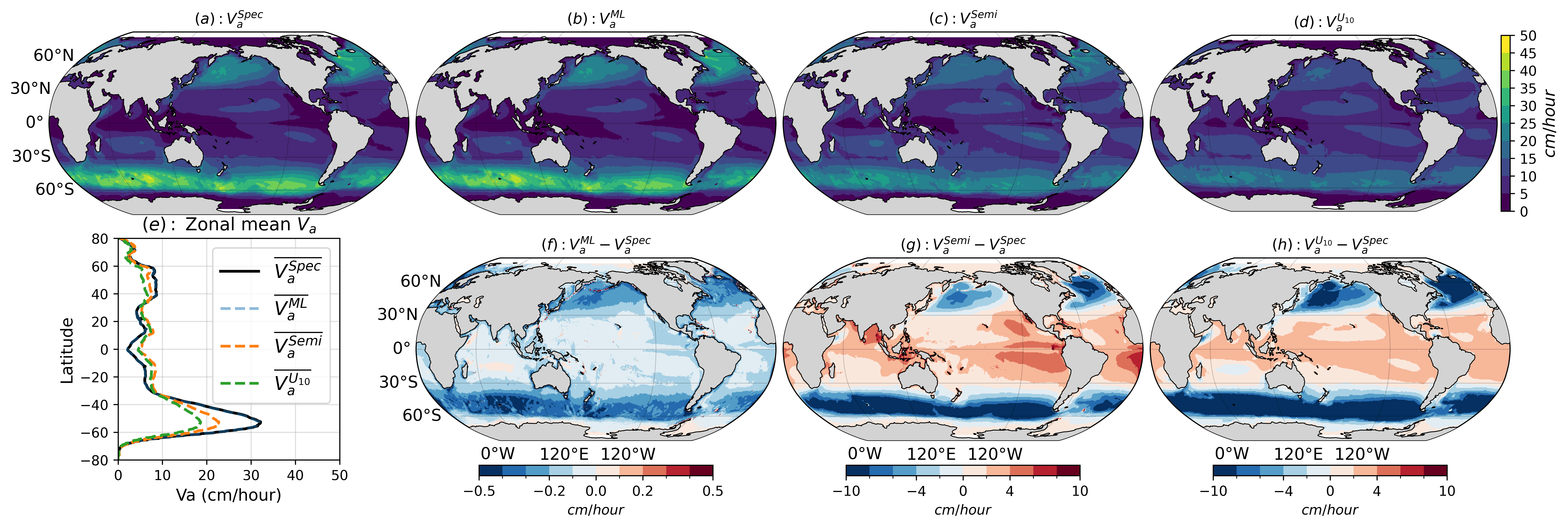}
    \caption{Annual mean air-entrainment velocity $V_a$ in $cm/hour$ from 2019–2022 for different parameterizations. Panels (a–d) show $V_a$ computed from (a) the spectral model $V_a^{Spec}$, (b) the machine-learning model $V_a^{ML}$, (c) the semi-empirical formulation $V_a^{Semi}$, and (d) the wind-speed-based parameterization $V_a^{U_{10}}$. Panel (e) presents the zonal-mean profiles for each scheme. Panels (f–h) display spatial differences relative to the spectral reference $V_a^{Spec}$. The machine-learning parameterization reproduces the large-scale structure and magnitude of the spectral model while substantially reducing biases in mid-latitude storm tracks and swell-dominated regions.}
    \label{fig:comparision}
\end{figure}

The machine learning driven predicted $V_a^{ML}$ shows notable improvement compared to the commonly used parameterization based on wind speed only ($V_a^{U_{10}}$) and a combination of frictional velocity and significant wave height ($V_a^{Semi}$) (Figure \ref{fig:comparision}). Figure \ref{fig:comparision} compares the global maps of annual mean $V_a$ for 2019–2022 derived from the three parameterizations. The reference spectral estimate $V_a^{Spec}$ (Figure~\ref{fig:comparision}{a}) has pronounced latitudinal variability, with maxima exceeding 40 cm h$^{-1}$ in the storm-track regions of the Southern Ocean and North Atlantic, and minima in the subtropical gyres. The machine-learning prediction $V_a^{ML}$ (Figure~\ref{fig:comparision}{b}) closely reproduces the global spatial pattern and magnitude of $V_a^{Spec}$, capturing enhanced entrainment under high-wind, young-sea conditions. The semi-empirical $V_a^{Semi}$ and wind-only $V_a^{U_{10}}$ formulations (Figure~\ref{fig:comparision}{c,d}) show similar large-scale structure but with reduced sensitivity to regional sea-state variability. The zonal-mean comparison (Figure~\ref{fig:comparision}{e}) demonstrates that $V_a^{\mathrm{ML}}$ aligns best with the spectral reference, while the bulk schemes underestimate entrainment at mid to high latitudes. Difference maps (Figure~\ref{fig:comparision}{f-h}) highlight systematic biases: the ML model shows tiny residuals ($<0.05$ cm/h$^{-1}$), whereas the semi-bulk and wind-based formulations underestimate $V_a$ in storm tracks and high latitudes, while overestimate it in swell-dominated tropical regions. Overall, the ML parameterization provides a faithful and computationally efficient surrogate for the spectral model, capturing both the global distribution and zonal structure of wave-breaking-induced air entrainment. 

Statistically, the machine-learning parameterization reproduces the spectral reference $V_a$ with a small negative bias ($-0.018$~cm~h$^{-1}$), $RMSE = 0.11$~cm~h$^{-1}$, $NRMSE = 0.08$ (normalized by the mean absolute reference), and an exceptionally high spatial correlation ($R = 0.999$) over 2019-2022. In contrast, the semi-empirical formulation ($V_a^{Semi}$) and the wind-only formulation ($V_a^{U_{10}}$) exhibit larger systematic underestimation (biases of $-0.081$ and $-0.279$~cm~h$^{-1}$, respectively), substantially larger errors ($RMSE = 1.14$ and $1.51$~cm~h$^{-1}$; $NRMSE = 0.78$ and $1.04$), and lower correlations ($R = 0.87$ and $0.92$). These results demonstrate that the ML approach captures the nonlinear dependence of air-entrainment velocity on wind forcing and sea state, substantially improving upon traditional empirical parameterizations.


\subsection{Validation against independent observations}
In addition to the 2019–2022 global simulations, the machine-learning model is evaluated against an independent observational dataset from the HiWinGS field campaign \cite{Brumer2017}, conducted aboard the R/V \textit{Knorr} in the North Atlantic. Hourly wave statistics during HiWinGS were derived from a combination of directional wave spectra measured by the Datawell Waverider buoy and WAMOS, supplemented by a cruise-specific WAVEWATCH III (WW3) hindcast, as well as one-dimensional spectra obtained from a Riegl laser altimeter following \citeA{Brumer2017}. For each observation, the breaker-front distribution $\Lambda(c)$ was estimated from the measured wave statistics and reconstructed wave spectra using the framework of \citeA{Phillips1985} and \citeA{Romero2012}. The corresponding air-entrainment velocity $V_a$ was then computed following \citeA{Deike2017} and \citeA{zhou2023sea}.

\begin{figure}[!htbp]
    \centering
    \includegraphics[width=0.6\linewidth]{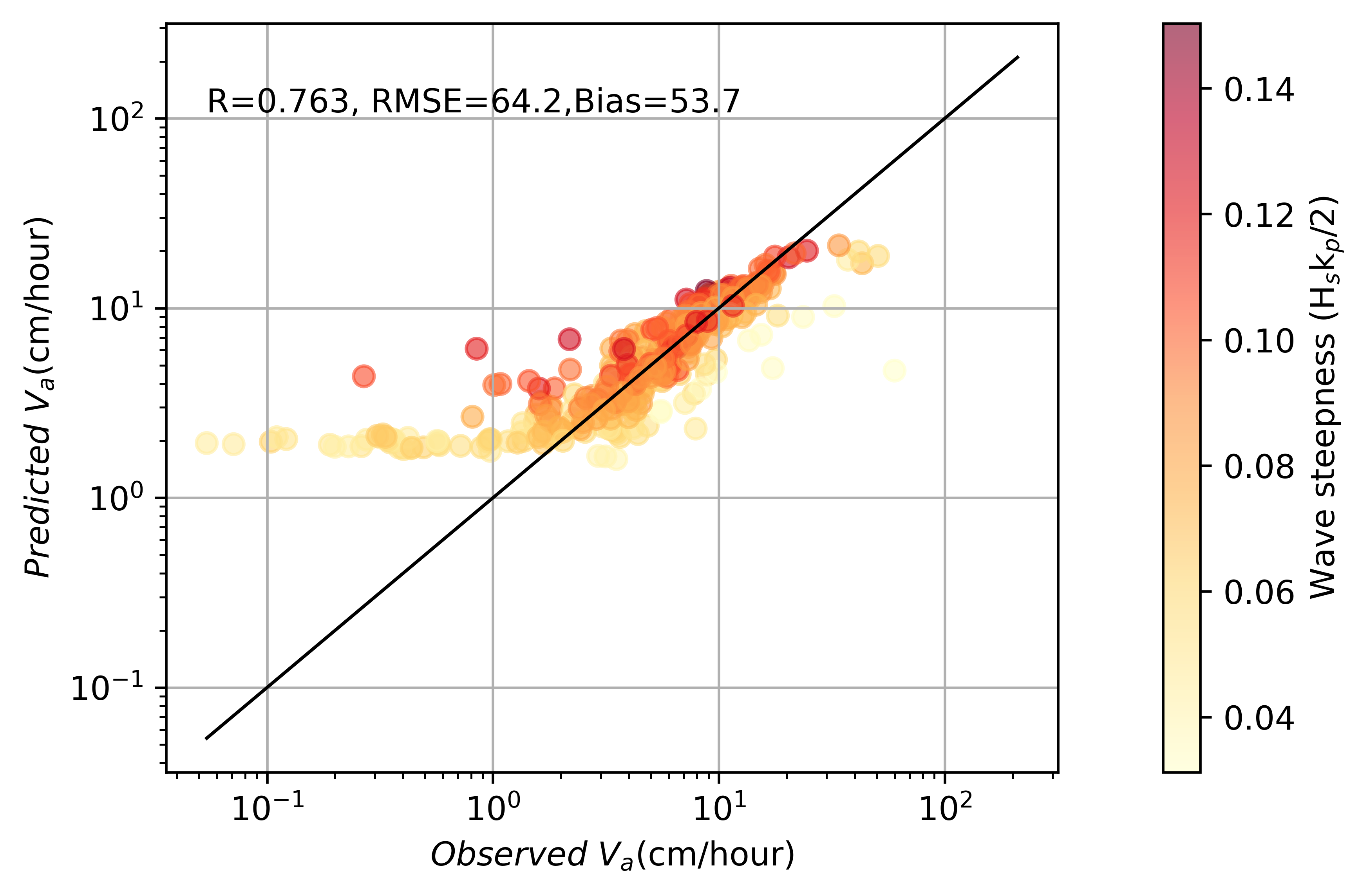}
    \caption{Comparison between predicted and observed air-entrainment velocity ($V_a$) from the HiWinGS field campaign \cite{Brumer2017}. Colors indicate wave steepness ($H_s k_p/2$).}
    \label{fig:observations}
\end{figure}

The trained model was applied to the observed sea-state variables—including significant wave height, 10-m wind speed and direction, wave age, wave steepness, and water depth—to infer $V_a$ during the HiWinGS cruise. Figure~\ref{fig:observations} compares predicted and observed air-entrainment velocities, with symbol color indicating wave steepness $H_s k_p / 2$. The machine-learning scheme reproduces the observed variability with a correlation coefficient of $R=0.76$, an RMSE of $64.2$~cm~hour$^{-1}$, and a bias of $53.7$~cm~hour$^{-1}$, demonstrating skill for an unseen observational dataset. Predictive performance is highest under wind-driven, steep-wave conditions, whereas the model tends to overestimate air entrainment under low-wind conditions and weak wave breaking (Figure~\ref{fig:observations}). In contrast, when applied to low-to-moderate wind conditions from ASIT \cite{hogan2025observations}(not shown), the model does not reproduce observed $V_a$. This likely reflects non-equilibrium conditions (fetch-limited and intermittent breaking) that violate the stationarity assumptions underlying the Phillips (1985) $\Lambda(c)$ reconstruction, plus low-$c$ detection limits and potential under-resolution of weak microbreaking. These factors weaken the link between spectral saturation and $\Lambda(c)$, so integral formulations can misrepresent $V_a$ outside the high-wind, near-equilibrium regime sampled by HiWinGS.

\subsection{Application for gas flux and sea salt emission}
Since wave breaking induced bubbles impact the air-sea gas transfer velocity \cite{Deike2017,zhou2023sea} and the sea salt emission \cite{deike2022mechanistic}. We can evaluate the machine learning model  presented in section 2.2 for the global application for wave breaking induced gas transfer velocity $k_b$ and sea salt emission due to wave breaking $M_{salt}$.

\subsubsection{Global bubble mediated gas transfer velocity}
Because breaking-driven bubbles influence both gas transfer velocity and sea-salt aerosol production \cite{Deike2017,zhou2023sea,deike2022mass}, we assess how the ML-based $V_a$ affects global estimates of bubble-mediated gas transfer velocity ($k_b$) and sea-salt emission.
Following \citeA{zhou2023sea}, the wave breaking induced bubbles generated gas transfer velocity $k_b$ can be parameterized as
\begin{equation}
    k_b = \frac{1}{2\pi \alpha}\frac{V_a}{C_A}\int \frac{4\pi R_b^3}{3} q(R_b)E(R_b) \, dR_b \, ,
\end{equation}
where $\alpha = K_oRT$ is the Ostwald dimensionless solubility coefficient, $K_o$ ($mol/m^3/atm$) is the gas solubility in  sea water, $R$ is the ideal gas constant ($J/K/mol$), and $T$ is the sea surface skin temperature with unit $K$. The size distribution of bubbles entrained under a breaking wave, $q(Rb)$, is defined as the number of bubbles per bin size, per unit volume, follows \cite{deane2002scale,deike2022mechanistic,mostert2022high}
\begin{equation}
    q(R_b) \propto
        \begin{cases}
        R_b^{-10/3}, & R_b > R_H, \\
        R_b^{-3/2},  & R_b < R_H .
\end{cases}
\end{equation}
In addition, $E(r)$ is defined as a function of the depth of bubble injection $z_0$ and an equilibration depth $H_{eq}$ together with the gas diffusivity and solubility. The equilibrium depth $H_{eq}$ is a function of the bubble rise velocity $w_b(r)$ and the exchange velocity for a single bubble $\kappa_b(r)$, which depends on bubble radius $r$, solubility and gas diffusivity of gas in sea water. Here we use the bubble rise velocity $w_b(r)$ in quiescent water following \citeA{keeling1993role} and \citeA{deike2018gas}; the details are described in \citeA{zhou2023sea}. $C_A$ is used to insure the volume is constrained by the third moment and dimensional consistency. In Figure \ref{fig:k_b}, panel (a) shows the global zonal and monthly mean bubble-mediated gas transfer velocity $k_b$ with increasing latitude from the spectrum simulation $k_b^{Spec}$, with unit $cm/hour$. The wave breaking induced gas transfer velocity are significant at high latitude and stronger in Boreal Winter and early spring, which is consistent with \cite{zhou2023sea, reichl2020contribution}. Panels (b) to (d) in Figure \ref{fig:k_b} show the difference of zonal monthly mean bubble mediated gas transfer velocity from machine learning model and the spectrum model $k_b^{ML}-k_b^{Spec}$, the difference of $k_b^{Semi}$ based on significant wave height based air entrainment reproduced gas transfer velocity and $k_b^{Spec}$, and the difference of gas transfer velocity from air entrainment based wind speed only $k_b^{U_{10}}$ and $k_b^{Spec}$. The bias between the  machine learning model and the spectrum model is in the rage of $0.1$ to $0.4$ $cm/hour$ or about 1\% in terms of relative difference. This is a significant reduction $O(10)$ when compared with traditional significant wave height and  wind speed dependent only parameterizations.

The bubble mediated gas transfer velocity computed with the  simplified parameterizations of $V_a^{U_10}$ and $V_a^{Semi}$ is overestimated in both tropical and subtropical regions due to significant wave swell signal and underestimated in the high latitude region, especially in winter seasons. The relative bias can approach $\sim 10\%$ of the gas transfer velocity. The $10\%$ bias of gas transfer velocity translates to a global bias of around $15\sim 20 PgC$ in oceanic carbon storage. The machine learning model in this study outperforms commonly used parameterizations while also being computationally efficient, qualities that are important for Earth system models.

\subsubsection{Global sea salt emission}
\citeA{deike2022mechanistic} proposed a sea state dependent sea spray generation function (SSGF) based on wave breaking and bubble bursting from two mechanisms (1) film bursting leading to film drops, and (2) cavity collapse leading to jet drops as following the previous studies \cite{lewis2004sea,veron2012sea}. The wave breaking induced SSGF is: 
\begin{equation}\label{eq3}
F_d(r_d) = \frac{1}{2\pi} \frac{V_a}{C_A}
\int \frac{q(R_b)\,n(R_b)}{\langle r_d \rangle(R_b)}\,
p\!\left(\frac{r_d}{\langle r_d \rangle}\right)\, dR_b
\end{equation}
where the $n(Rb)$, $<rd>$ are respectively the number and mean radius of drops resulting from one of the mode of production (jet and film drops), the size distribution $q(Rb)$ of bubbles entrained under a breaking waves, and defined as the number of bubbles per bin size, per unit volume, follows $q(R_b) \propto R_b^{-10/3}, \quad R_b > R_H$, while $q(R_b) \propto R_b^{-3/2}, \quad R_b < R_H$\cite{deane2002scale,deike2022mechanistic,mostert2022high}. 

The distribution of drop size production by bubble bursting is then be modeled by a Gamma distribution $P$ \cite{villermaux2020fragmentation}:
\begin{displaymath}
    P(r_d/<r_d>) = \frac{m^m}{\Gamma(m)}(\frac{r_d}{<r_d>})^{m-1} e^{-m\frac{r_d}{<r_d>}}
\end{displaymath} with $m = 11$.
The bubble radius in this study ranges from $60 \mu m \sim 1.2 mm$ following \citeA{Deike2017}.

The mechanisms controlling the number and mean droplet radius produced by bubble bursting are described in detail by \citeA{deike2022mechanistic}; here we summarize the three modes used in this study. Jet drops form when the bubble cavity collapses and produces an upward jet that breaks into droplets. Jet drops are generated by bubbles with radius $R_b<l_{\mu}$, with $n_{\rm jet}(R_b)=\chi_{1}(R_b/l_{\mu})^{-1/3}$ and $\langle r_d\rangle_{\rm jet}=\chi_{2}l_{\mu}(R_b/l_{\mu})^{5/4}$, where $\chi_1=145$, $\chi_2=0.008$, and $l_{\mu}$ is the visco-capillary length.

Film drops arise from rupture of the thin liquid film separating the bubble from the atmosphere and occur in two modes \citeA{deike2022mechanistic}. Mode I (centrifugal ejection) applies to $R_b\approx0.6$--$10$~mm in seawater, with $n_{\rm filmI}(R_b)=\chi_{3}(R_b/l_{c})^{2}(R_b/h_b)^{7/8}$ and $\langle r_d\rangle_{\rm filmI}=\chi_{4}R_b^{3/8}h_b^{5/8}$, where $h_b=\chi_{5}Sc^{-1/2}R_b^2/l_c$ and $\chi_3=0.04$, $\chi_4=0.5$, $\chi_5=0.04$ \cite{villermaux2020fragmentation}. Mode II (flapping) applies to $R_b\approx100~\mu$m--$1.2$~mm, with $n_{\rm filmII}(R_b)=\chi_{6}Sc^{-1/2}(R_b/l_{c})^{2}$ and $\langle r_d\rangle_{\rm filmII}=\chi_{7}Sc^{-1/2}(R_b/l_{c})^{1/3}$, where $\chi_6=3.6\times10^6$ and $\chi_7=5\times10^{-5}$ \cite{jiang2022submicron}. The total sea-spray generation function is the sum of these three contributions.

 The emission of sea salt is estimated following \citeA{de2011production} and \citeA{lewis2004sea}. The  mass of a single particle is $m_{dry} = \rho_{ss} 4/3\pi (r_d^{80})^3$, where $\rho_{ss}=1.22 g/cm^3$ is the sea salt density. Based on the sea salt generation function from \citeA{deike2022mechanistic}, we estimated the global sea salt emission with the diameter $D<10 \mu m$ from the wave spectrum based air entrainment $V_a$, the machine learning model reproduced $V_a^{ML}$, the significant wave height reproduced $V_a^{Semi}$ and the wind speed only parameterized $V_a^{U_{10}}$. 
 
\begin{figure}[!htbp]
    \centering
    \includegraphics[width=\linewidth]{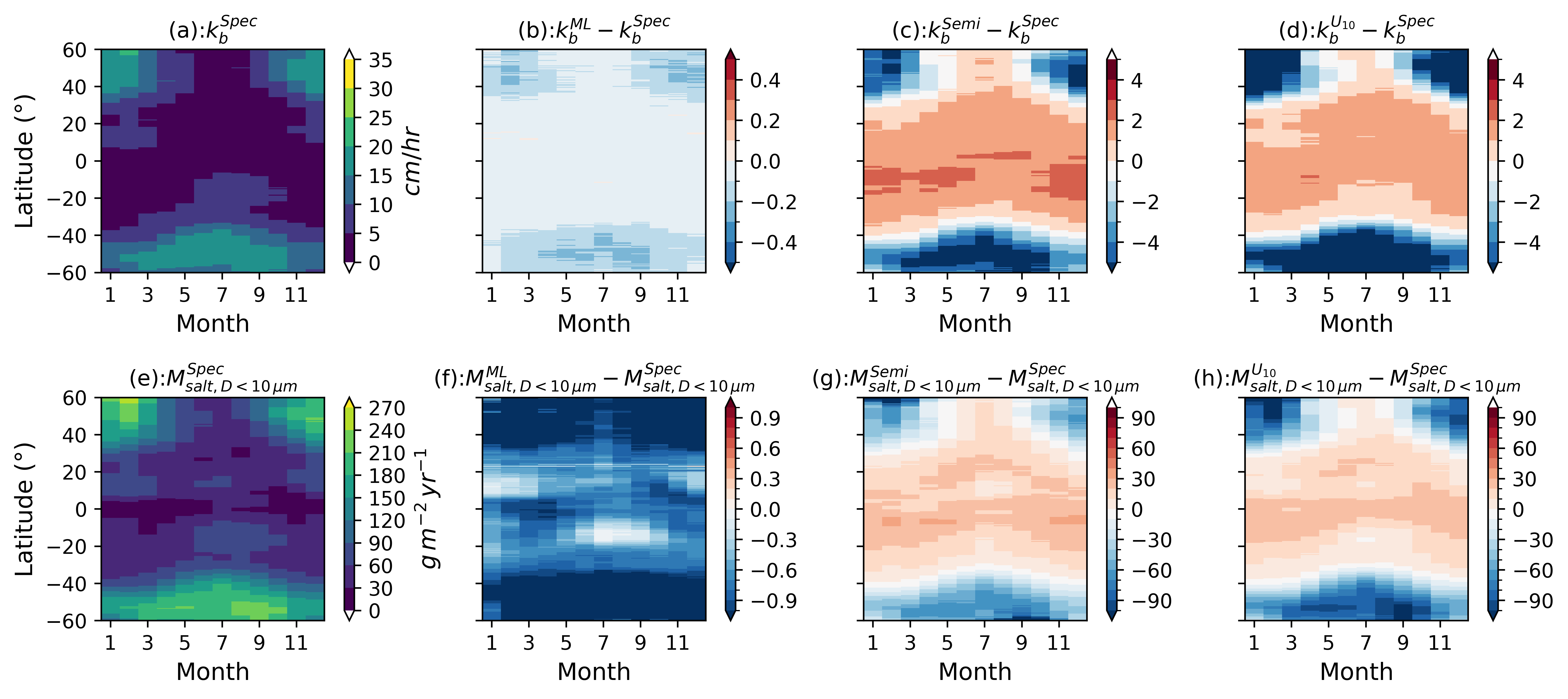}
    \caption{a) The global zonal monthly bubble mediated gas transfer velocity from the wave spectrum model predicted air entrainment velocity $k_b^{spec}$ from 2019 to 2022.  b) The difference of bubble mediated gas transfer velocity from air entrainment velocity based on different model, the difference between machine learning model and wave spectrum model. c) The difference between semi-bulk model based on significant wave height and wave spectrum model. d) The difference between bulk model based on wind speed only and wave spectrum model. Panels (e) to (h) similar to the previous panels, but for the sea salt emissions with droplet diameter less than $10 \mu m$.}
    \label{fig:k_b}
\end{figure}

 Figure \ref{fig:k_b} (e) shows the zonal monthly mean sea salt emission for $D_p < 10 \mu m$ from the wave spectrum model simulated $V_a^{Spec}$. The global pattern of sea salt emissions is similar to gas transfer velocity. This is expected, as both quantities are directly related to wave breaking. The estimated global emission with $D<10 \mu m$ are about $22 Tg/yr$. The machine learning model can be nearly identical with the spectrum model simulated global sea salt emission as shown in Figure \ref{fig:k_b} (f). However, both significant wave height and wind speed dependent parameterization introduce a nearly $30\%$ bias for sea salt emission. They overestimate the emission in tropical and underestimate it at high latitude. The mechanism of such significant bias is due to the strong swell in tropical and subtropical, while the strong swell don't directly produce wave breaking as discussed in \citeA{zhou2023sea}. 
\section{Conclusions and discussion }
We developed an MLP-based parameterization of wave-breaking air-entrainment velocity $V_a$, a key quantity controlling bubble-mediated air-sea exchange. The work is motivated by the computational cost of diagnosing $V_a$ from a full spectral model with $\Lambda(c)$, and by systematic regional biases in commonly used wind-only and semi-bulk formulations. The ML model reproduces the spectral-reference $V_a$ globally and shows skill against independent HiWinGS observations. When propagated into applications, the ML-based $V_a$ substantially reduces regional biases in bubble-mediated gas transfer velocity and sea-salt emissions relative to traditional parameterizations.

This framework is currently limited by the Phillips (1985) equilibrium-breaking assumption and is most reliable under deep-water, near-equilibrium conditions. As additional observations become available—particularly in shallow-water and non-equilibrium regimes—the approach can be extended by retraining with expanded training targets. Overall, the parameterization provides an efficient pathway to improve sea-state-dependent air-sea flux representations in Earth system models.

\section*{Open Research Section}
WAVEWATCH~III is publicly available (\url{https://github.com/NOAA-EMC/WW3}). The implementation of the breaking-crest distribution $\Lambda(c)$ in WWIII, following \cite{Romero2019}, is available at \url{https://github.com/Leonel-Romero/WW3-Lambda}. For peer review, the wave-spectral simulation outputs used in this study, the machine-learning predictions, and the trained machine-learning model (weights and configuration) are available at \url{https://utdallas.app.box.com/v/zhou-jgr-2026}. HiWinGS cruise data products used for validation are publicly available via the NOAA PSL cruise archive \url{https://psl.noaa.gov/data/cruises/}.

\section*{Conflict of Interest declaration}
The authors declare there are no conflicts of interest for this manuscript.

\acknowledgments
This research was supported by the NASA Modeling, Analysis, and Prediction (MAP) Program through the Global Modeling and Assimilation Office: Core Activities, and by the National Science Foundation (NSF) under grant number 2319535. Any opinions, findings, and conclusions or recommendations expressed in this material are those of the authors and do not necessarily reflect the views of NSF. 
The computational resources were provided by NOAA’s RDHPC systems, Texas Advanced Computing Center (TACC) at the University of Texas at Austin through allocations OCE24006 and OCE24007, and Purdue Anvil CPU at Purdue University through allocation PHY240204 from the Advanced Cyberinfrastructure Coordination Ecosystem: Services \& Support (ACCESS) program.

%
\bibliography{references} 
%


%
%
%
%
%

\end{document}